# Polarized deep diffractive neural network for classification, generation, multiplexing and de-multiplexing of orbital angular momentum modes

Jiaqi Zhang, [1, 2] Zhiyuan Ye, [3] Jianhua Yin, [1] Liying Lang, [4, *] and Shuming Jiao[1, *]

[1] Peng Cheng Laboratory, Shenzhen, Guangdong 518055, China
[2] Center for Terahertz Waves and College of Precision Instrument and Optoelectronics Engineering, and Key Laboratory of Optoelectronic Information Technology (Ministry of Education), Tianjin University, Tianjin 300072, China
[3] Department of Physics, Applied Optics Beijing Area Major Laboratory, Beijing Normal University, Beijing 100875, China
[4] Center for Advanced Laser Technology and Hebei Key Laboratory of Advanced Laser Technology and Equipment and Tianjin Key Laboratory of Electronic Materials and Devices, Hebei University of Technology, Tianjin 300401, China

*Corresponding author: jiaoshm@pcl.ac.cn;  langliying@hebut.edu.cn

## Abstract

The multiplexing and de-multiplexing of orbital angular momentum (OAM) beams are critical issues in optical communication. Optical diffractive neural networks have been introduced to perform classification, generation, multiplexing and de-multiplexing of OAM beams. However, conventional diffractive neural networks cannot handle OAM modes with a varying spatial distribution of polarization directions. Herein, we propose a polarized optical deep diffractive neural network that is designed based on the concept of rectangular micro-structure meta-material. Our proposed polarized optical diffractive neural network is trained to classify, generate, multiplex and de-multiplex polarized OAM beams. The simulation results show that our network framework can successfully classify 14 kinds of orthogonally polarized vortex beams and de-multiplex the hybrid OAM beams into Gauss beams at two, three and four spatial positions respectively. 6 polarized OAM beams with identical total intensity and 8 cylinder vector beams with different topology charges also have been classified effectively. Additionally, results reveal that the network can generate hybrid OAM beams with high quality and multiplex two polarized linear beams into 8 kinds of cylinder vector beams.

## 1 Introduction

Vortex beams with helical phase front, possessing orbital angular momentum (OAM) or topological charge ($l$), have attracted wide attention in many pivotal fields such as optical communication[1,2], optical tweezers[3,4], quantum information processing[5,6] and holographic display[7,8], etc. The ring radius of vortex beam is a straightforward feature for the identification of topological charge number[9]. As a critical technology in optical communication, vortex beams with different modes or topological charge numbers have been considered as multiple independent communication channels to improve communication bandwidth and capacity[10-12]. For example, 16 OAM beams with two polarizations and a single wavelength can achieve 2.56 Tbit/s free space communication[11]. This mainly takes the advantages of multiplexing of the OAM modes. Besides, there are also other multiplexing technologies combined with the OAM mode multiplexing to empower optical communication efficiency, such as wavelength-division multiplexing[13,14], space-division multiplexing[15,16], polarization-division multiplexing[17,18] and time-division multiplexing[19,20]. Various methods have been studied to generate and modulate vortex beams with different OAM modes and polarization states such as spiral phase plate[21,22], liquid crystal spatial light modulator[23], micro-structure meta-material[24,25] and deep diffractive neural network ($D^2NN$)[26,27]. Some researchers generate a specific cylinder vector beam by the interference of vortex beams[28]. \par
Deep learning as the most widely used method in artificial intelligence has made a great advance in many tasks such as image classification[29], speech recognition[30], chemical component analysis[31] and language translation[32] etc. Optical neural network (ONN) is a combination between optics and deep learning which was firstly proposed as an optical vector-matrix multiplication model by J.W. Goodman of Stanford university in 1978[33]. $D^2NN$ consisting of multiple diffractive layers is another typical physical implementation of ONN. $D^2NN$ has been widely investigated for various optical computing tasks including number digit classification[34-36], optical logic gate operation[37] and multiplexing and de-multiplexing of OAM beams[26]. For example, Z. Huang et al. designed a $D^2NN$ to achieve a depth-controllable imaging technology in OAM multiplexing holography[26,38]. For improving the accuracy of $D^2NN$, many nonlinear materials and nonlinear physical processes designed as nonlinear optical activation functions have been introduced into $D^2NN$ [39]. Additionally, in order to handle different machine learning tasks for wave sensing, programmable meta-surfaces have also been proposed to achieve in-situ modulation of diffractive layers[40].

However, in most cases, the $D^2NN$ cannot perform light modulation in the dimension of polarization. Namely, there is no polarization anisotropy in most previously proposed $D^2NN$s. Recently meta-surface has been widely reported to achieve anisotropic polarized light modulation[41,42]. For example, the rectangular microstructure based on the concepts of Huygens meta-surface and Berry phase is the common anisotropic meta-surface, which is favorable for us to design optical device with polarization anisotropy. In this paper, we propose a polarized deep diffractive neural network ($PD^2NN$) to classify, generate, multiplex and de-multiplex polarized OAM beams, which is modulated by rectangular pillar meta-materials. The proposed $PD^2NN$ consists of multiple anisotropic pure phase layers which are optimized by wave front matching algorithm[43]. The results show our $PD^2NN$ can accurately recognize the trained vortex beams or cylinder vector beams. In addition, our proposed network can spatially de-multiplex hybrid polarized OAM beams into spatially separated Gauss beams. From the aspect of generation and multiplexing, our $PD^2NN$ can achieve the objective of generating two kinds of high quality cylinder vector beams and multiplexing two different polarized Gauss beams into 8 kinds of cylinder vector beams with multiple polarized modes.

## 2 Model and principles of $PD^2NN$

### 2.1 Matrix model of polarized light modulation by an anisotropic neural layer

Various materials and methods proposed previously can be used as anisotropic neural layer in a $D^2NN$. Here, we consider taking rectangular meta-material to design our anisotropic neural layers in which x-polarized and y-polarized have different responses in horizontal and vertical directions. In a common rectangular meta-material structure, the long and short sides correspond to the modulation of a-polarized or b-polarized incidence. And anisotropic amplitude and phase modulation can be achieved by rotating the angle of rectangular micro-structure, the structure diagrams of rectangular meta-material are shown in Fig. 1. The period of the micro pillar is P, the length and width of the rectangular pillar are a and b respectively and the angle $\theta$ is the angle between long side and x axis.

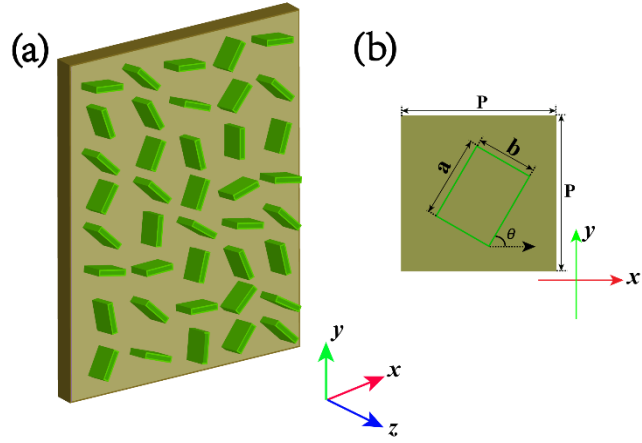

Figure 1. (a) The structure diagram of rectangular pillar micro-structure. (b) The rectangular pillar unit cell.

In general, polarized light can be expressed by Jones vectors and the electric field in x, y coordinates can be expressed as a column vector: $E = [E_x\ E_y]^T$ where $E_x$ and $E_y$ correspond to x-polarized and y-polarized components. The transmission matrix $T_s$ of a single-pillar unit cell in a and b coordinate can be represented as:

$$T_s = \begin{bmatrix} t_a e^{i\varphi_a} & 0 \\ 0 & t_b e^{i\varphi_b} \end{bmatrix}$$

Where $t_a$, $t_b$, $\varphi_a$ and $\varphi_b$ are the transmission coefficient of amplitude of a-polarized and b-polarized incidence and the transmission coefficient of phase of a-polarized and b-polarization respectively. In our calculation, $t_a=t_b=1$ which are determined by anisotropic amplitude modulation effect. In other words, our designed structure has no modulation in amplitude but only in phase. The transmission matrix T can be obtained by Eq. 1 when (a, b) coordinates have an in-plane anti-clockwise rotation angle $\theta$ with respect to the (x, y) coordinates as shown in Fig. 1. We set $\varphi = \varphi_b - \varphi_a$, and the transmission matrix $T$ can be:

$$T = \begin{bmatrix} \cos\theta & -\sin\theta \\ \sin\theta & \cos\theta \end{bmatrix} \begin{bmatrix} e^{-i\frac{\varphi}{2}} & 0 \\ 0 & e^{i\frac{\varphi}{2}} \end{bmatrix} \begin{bmatrix} \cos\theta & \sin\theta \\ -\sin\theta & \cos\theta \end{bmatrix}$$

Then the final transmission $R$ in x and y coordinate can be obtained by Eq.2:

$$R = TE = \begin{bmatrix} \cos\theta & -\sin\theta \\ \sin\theta & \cos\theta \end{bmatrix} \begin{bmatrix} e^{-i\frac{\varphi}{2}} & 0 \\ 0 & e^{i\frac{\varphi}{2}} \end{bmatrix} \begin{bmatrix} \cos\theta & \sin\theta \\ -\sin\theta & \cos\theta \end{bmatrix} \begin{bmatrix} E_x \\ E_y \end{bmatrix}$$

## 2.2 Free-space propagation model and optimization of neural network

Our proposed PD²NN structure has several hidden layers which have 256×256 neural units in every layer. Each neural unit has a size of 5 μm. The distance d between each hidden layer is 50 mm. The studied wavelength ($\lambda$) is set as 1500 nm. The network is trained by input vortex incidence or spatially separated Gauss beams which is depending on the specific task of network. To achieve a fully connected neural network, we firstly make input vortex incidence with 256×256 pixels. The structure diagram of PD²NN is shown in Figure 2.

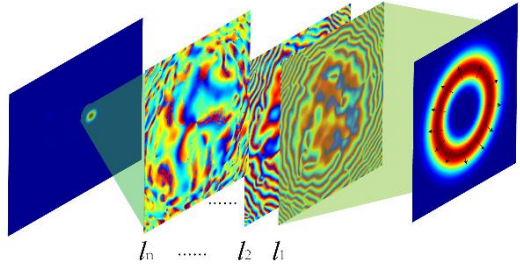

Figure 2. The structure diagram of proposed PD$^2$NN

The light field propagation theory of PD$^2$NN is simulated based on Rayleigh Sommerfeld diffractive theory. Every single neural unit of a given PD$^2$NN layer is regarded as a point source of a wave and the connection weight can be expressed as:

$$\omega_i^l(x,y,z) = \frac{z - z_i}{r^2}\left(\frac{1}{2\pi r} + \frac{1}{j\lambda}\right)\exp\left(\frac{j2\pi r}{\lambda}\right)$$

Where $l$ represents the $l$-th layer of the network, $i$ represents $i$-th neural unit which is located at (x, y, z) in hidden layer, $\lambda$ is the illumination wavelength, $r = \sqrt{(x-x_i)^2 + (y-y_i)^2 + (z-z_i)^2}$ which represents the distance of neural nodes between two adjacent hidden layers and $j = \sqrt{-1}$. We use wave-front matching algorithm to optimize the anisotropic phase information of hidden layers. According to wave-front matching algorithm, when the overall error between the modulated forward propagating fields and backwards propagating fields is minimum, the optimized phase can be obtained[43]:

$$\exp(j\varphi) = \text{Phase}\left[\sum_{n=1}^{N} E_{bn}\text{conj}(E_{fn})\right]$$

Where $\text{Phase}[\cdots]$ is operator of obtaining the phase of the electric field, $\text{conj}(\cdots)$ represents operator of phase conjugation, $\varphi$ is final optimized phase, $E_{bn}$ and $E_{fn}$ are backward propagation electric field in $n$-th layer and forward propagation electric field in $n$-th layer and $j = \sqrt{-1}$. Combining with anisotropic character, the matrix of final optimized phase can be expressed as:

$$\begin{bmatrix}\exp(j\varphi_x)\\ \exp(j\varphi_y)\end{bmatrix} = \begin{bmatrix}\text{Phase}[\sum_{n=1}^{N} E_{bn,x}\text{conj}(E_{fn,x})]\\ \text{Phase}[\sum_{n=1}^{N} E_{bn,y}\text{conj}(E_{fn,y})]\end{bmatrix}$$

Where $N$ represents the number of hidden diffractive layer, $n$ represents $n$-th layer, $\varphi_x$ is optimized phase in x axis, $\varphi_y$ is optimized phase in y axis, $E_{bn,x}$ and $E_{fn,x}$ are result of $E_{bn}$ and $E_{fn}$ along x axis and $E_{bn,y}$ and $E_{fn,y}$ are result of $E_{bn}$ and $E_{fn}$ along y axis. Finally, we can obtain $\theta$ and $\varphi$ and then we could design proper rectangular pillar with $\varphi_a$ and $\varphi_b$.

## 3 Classification, generation, multiplexing and de-multiplexing of polarized OAM beams with PD$^2$NN

### 3.1 Classification and de-multiplexing of polarized vortex beams with different topological charges

To demonstrate the performance of PD$^2$NN framework, we firstly trained 14 kinds of x polarized or y polarized vortex beam with topological charges $l$ = -3, -2, -1, 0, 1, 2, 3, respectively. We trained the network with 5 hidden layers which is phase-only modulated. In order to achieve the effect of classifying polarized vortex beams with different topological charges, we classify 14 kinds of polarized vortex beams with different topological charges into 14 Gauss beams in 14 different circle areas of detecting plane as shown in Fig. 3 (a). The Gauss beam in each circle area corresponds to a kind of polarized vortex beam. The intensity of x polarized incidence and y polarized incidence, phase of x polarized incidence and y polarized incidence and total intensity of x and y polarized incidence are shown in first, second, third, fourth and fifth rows of Fig. 3 (b) respectively. In our calculation, the polarized beams are expressed by two matrices. Taking x polarized beam for example, the sub-figures in the first row are beam spot intensity distributions and the subfigures in the second row denote zero intensities in y polarized direction as shown in Fig. 3 (b). The classification results are shown in the last row of Fig. 3 (b) indicated by total intensity of x and y polarized outputs in the detecting plane. Obviously, as shown in the first and second rows of Fig. 3 (b), the size of hollow spots becomes bigger with the absolute value of the topological charge of vortex beams increasing. Fig. 3 (d) is the zoom-in image of the classification result of x polarized vortex beam with $l = 1$ which indicates the output Gauss beam is located accurately in the pre-defined circle area. It can be observed that every Gauss output beam corresponds exactly to the target location in the detecting plane. These results verify the capability of the network with 5 hidden layers to classify polarized vortex beams with different topological charge numbers and polarized states. The trained phase $\varphi$ and rotation angle $\theta$ of 5 hidden layers are shown in first and second rows of figure 3 (c) respectively.

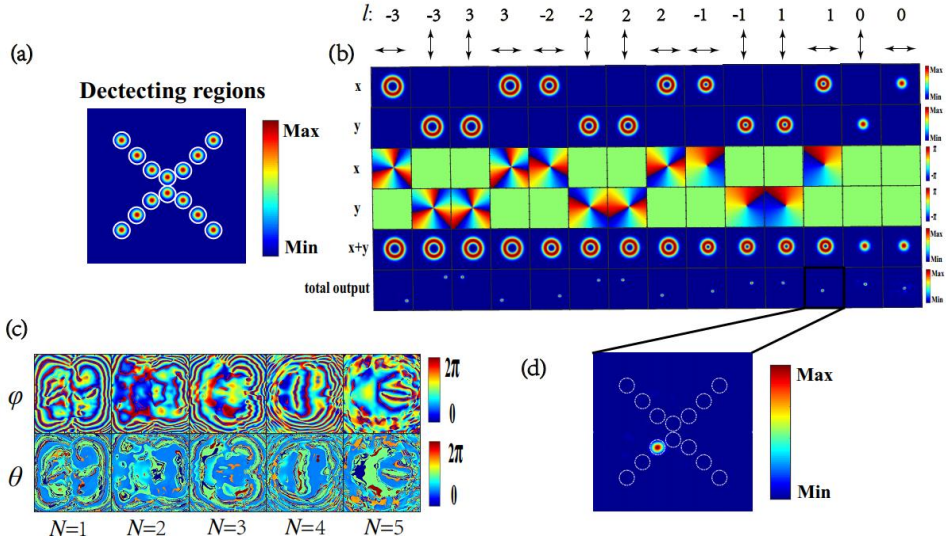

Figure 3. Classifying polarized vortex beams with different topological charges. (a) Ideal output plane with different classification results. (b) Incidence with different topological charge and the results of classification. (c) The trained phase and angle distributions of proposed PD²NN from first to fifth layer. (d) The zoom-in image of the output of multiplexing x polarized vortex beam with $l = 1$.

For quantitatively evaluating the quality of the classification results, we define signal to noise ratio (SNR) of target detecting regions to measure the accuracy of classification. The definition of the *SNR* can be seen as follows:

$$SNR_{i,j} = \frac{Signal_{i,j}}{Noise_i}$$

Where $SNR_{i,j}$ (i, j=1, 2, 3, …, 14) represents signal-to-noise ratio of the light intensity signal of *i*-th Gauss beam in *j*-th Gauss beam's target region, $Signal_{i,j}$ represents the light intensity of *i*-th

Gauss beam in *j*-th Gauss beam's target region, Noise$_i$ represents the light intensity of *i*-th Gauss beam in non-target area. The detecting regions in the output plane are divided into 14 circles as shown in Fig. 3 (a), the signal in a target region is only the total light intensity within a circle area and noise is the total intensity out of a circle area. The *SNR$ results of 14 detecting regions are shown in the heat map of Fig. 4. We can see clearly that the values of diagonal elements are much higher than the other values which indicates output Gauss beams mainly focus on the corresponding target regions.

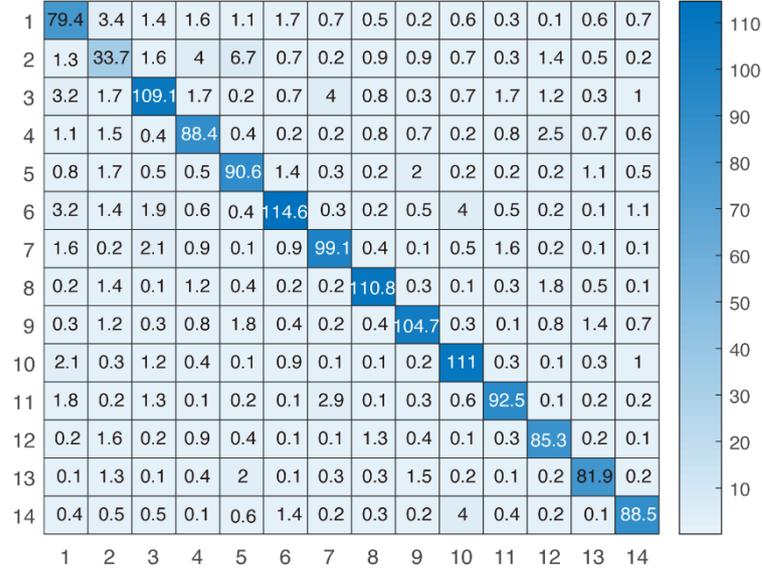

Figure 4. The heat map of signal to noise $SNR_{i,j}$ of detecting results for the classification of polarized vortex beams with different topological charges

The Fig. 5 shows that the proposed PD²NN network trained by 14 kinds of cross-polarized vortex beams with *l* = -3, -2, -1, 0, 1, 2, 3 respectively can spatially de-multiplex polarized hybrid OAM beams. In Fig. 5 (a), the first, second and third rows represent the de-multiplexing results of hybrid OAM modes with two (x polarized vortex beams with *l* = 2 and y polarized vortex beams with *l* = 3), three (x polarized vortex beams with *l* = 1, 3 and y polarized vortex beams with *l* = -3) and four (x polarized vortex beams with *l* = 1, -3 and y polarized vortex beams with *l* = -1, -2) kinds of modes respectively. We can see clearly that two, three and four corresponding Gauss beams in the output plane are generated. The *SNR* of spatially de-multiplexing two, three and four hybrid polarized OAM modes are shown in Figs. 5 (a), (b) and (c) respectively. The red bars in Figs. 5 (b), (c) and (d) represent the signal to noise of results of de-multiplexing hybrid polarized OAM modes in which the de-multiplexed signals can be distinguished from noise in non-target regions.

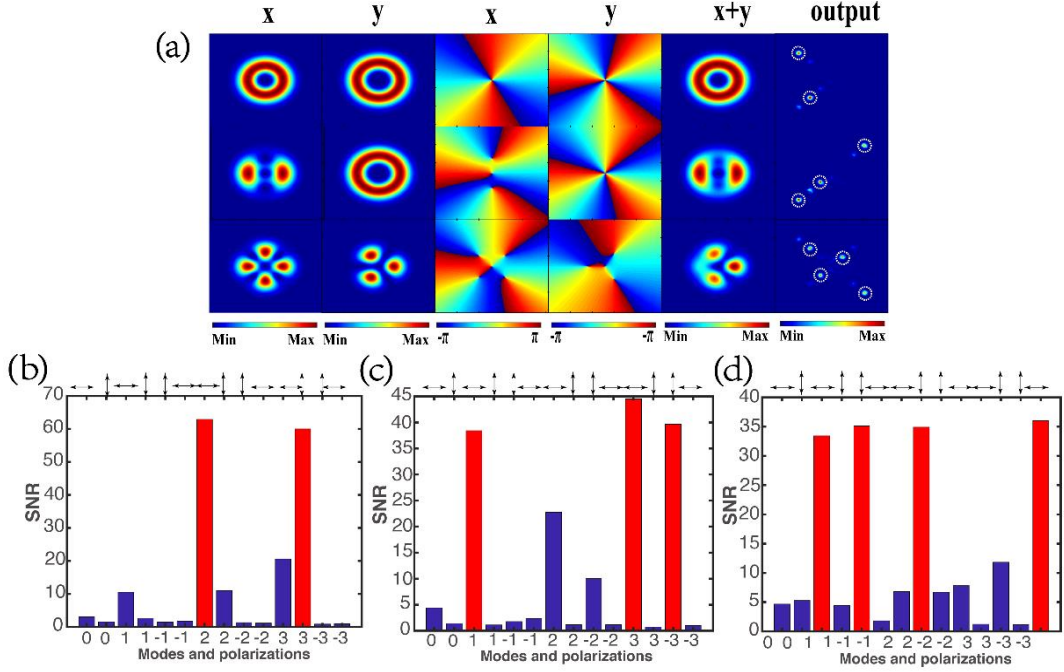

Figure 5. Spatial de-multiplexing of hybrid polarized OAM beams. (a) Input polarized beams with multiple topological charges and output results. The de-multiplexing *SNR* of polarized OAM beams summed by (b) two (c) three (d) four polarized Gauss beams. The red bars are corresponding to the results of de-multiplexing modes.

## 3.2 Classification of polarized OAM beams with identical total intensity

In order to verify the capability of our network to classify polarized OAM beams with identical total intensity, we trained our PD$^2$NN network with 5 hidden layers to classify four kinds of x or y polarized vortex beams with $l$ = -2 and 2 and two kinds of cylinder vector beams which have an identical total intensity in x and y polarization directions with the first four kinds of vortex beams. The intensity patterns of x and y polarized cylinder vector beams are obtained from interference of vortex beams with $l$ = -1 and 1. The patterns of detecting plane are shown in Figs. 6 (a) which is divided into six circle areas in different locations. We can see clearly that the total intensities of the six beams in the input plane in Fig. 6 (b) are identical which cannot be distinguished by human eyes while the intensities of each polarization direction are different. Fortunately, our network can convert the 6 polarized beams into 6 Gauss beams in the output plane at the target locations as shown in the last column of Fig. 6 (b). In the last two rows of the detecting plane in Fig. 6 (b), there are some noise signals in non-targeted locations. This is because the 6 polarized beams with identical total intensity are not orthogonal in topological states which will affect the classification performance. It is evident that the two kinds of cylinder vector beams can be interfered by the first four kinds of polarized vortex beams. In other words, the input cylinder vector beam may include modes of vortex beams in the front which makes the output plane exhibit other possible results. Although the noise signals are relatively stronger, we can still easily distinguish the target signal from the noise signal in the output plane. The final optimized rotation angle and modulation phase distributions of each diffractive layer are shown in Fig. 6 (c).

The $SNR_{i,j}$ ($i,j$=1, 2, 3, …, 6) of classification results of polarized beams with identical total intensity are given in Fig. 6 (d) where $i$ and $j$ represent light with different polarization states from 1 to 6 rows in Fig. 6 (d) respectively. The heat map tells us that there is high contrast between The

$SNR_{i,j}$ ($i, j$=1, 2, 3, 4) while there is low contrast between The $SNR_{i,j}$ ($i, j$=5, 6) which indicates some disturbing signals are emerging when detecting the results of multiplexing of two kinds of cylinder vector beams possessing the same total intensity with the first four vortex beams. Although the contrast between $SNR_{i,j}$ ($i, j$=5, 6) is low, the target signals in specific detecting regions still can be distinguished from noise signals in non-target detecting regions.

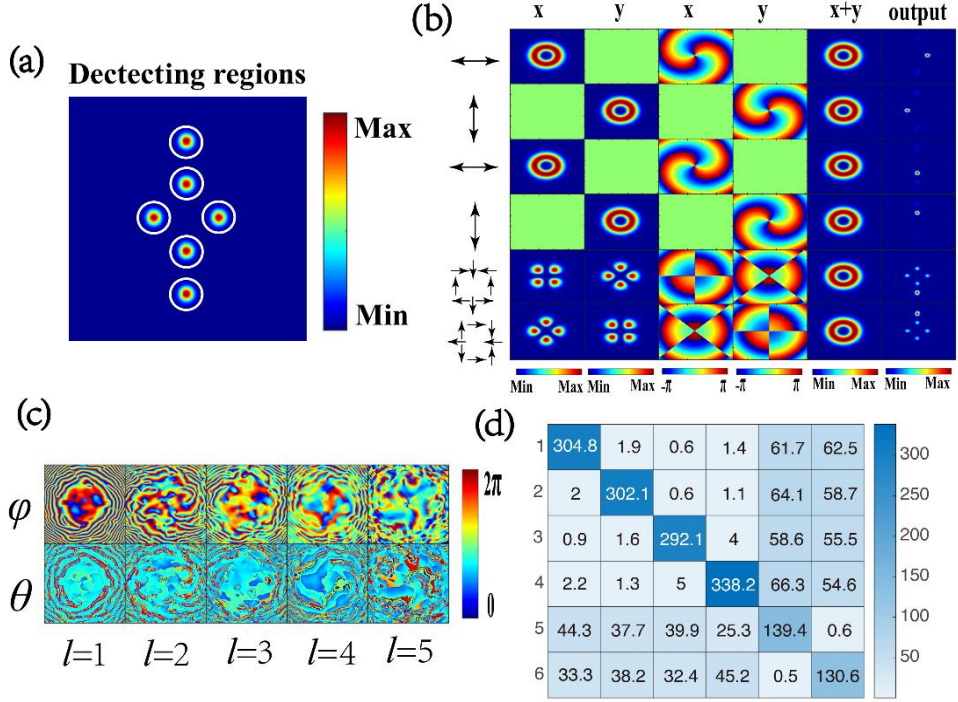

Figure 6. Classifying polarized vortex beams with identical topological charge. (a) Incidence with identical topological charge and the results of classifier. (b) The output plane with different classification results. (c) The optimized rotation angle and phase. (d) The heat map of signal to noise of classifying polarized beams with identical total intensity.

### 3.3 Classification of cylinder vector beams with different topological charge numbers

For classifying cylinder vector beams with different topological charge numbers and polarization states, we trained the PD²NN network with 5 layers to classify 8 kinds of cylinder vector beams which are generated by interference of vortex beams. Eight kinds of cylinder vector beams with different polarization states and topological charge numbers are assigned into 8 circle areas in output plane as shown in Fig. 7 (a). The radial and angular polarization beams of 1, 2, 3 and 4 orders are the interference results of vortex beams with $l = \pm1, \pm2, \pm3$ and $\pm4$ respectively. As shown in first and second columns of Fig. 7 (b), with the order of polarization beams increasing, the petals of the intensity patterns of cylinder vector beams are increasing. And the sizes of hollow spots which is the total intensity of x and y components are becoming larger as the order of cylinder vector beams increases as shown in Fig. 7 (b). The results shown in the last column of Fig. 7 (b) indicate that the total intensity of x and y polarized output Gauss beams are exactly located in the target detecting regions as shown in Fig. 6 (a). And the final optimized phase and rotation angle distributions are shown in Fig. 7 (c). The quantitative results of $SNR_{i,j}$ ($i, j$=1, 2, 3, …, 8) of classifying cylinder vector beams with different topological charge numbers are shown in Fig. 7 (d). In Fig. 7 (d), the minimum of contrast ratio between diagonal elements and other elements are larger than 22.6 which is sufficient to distinguish the target signal.

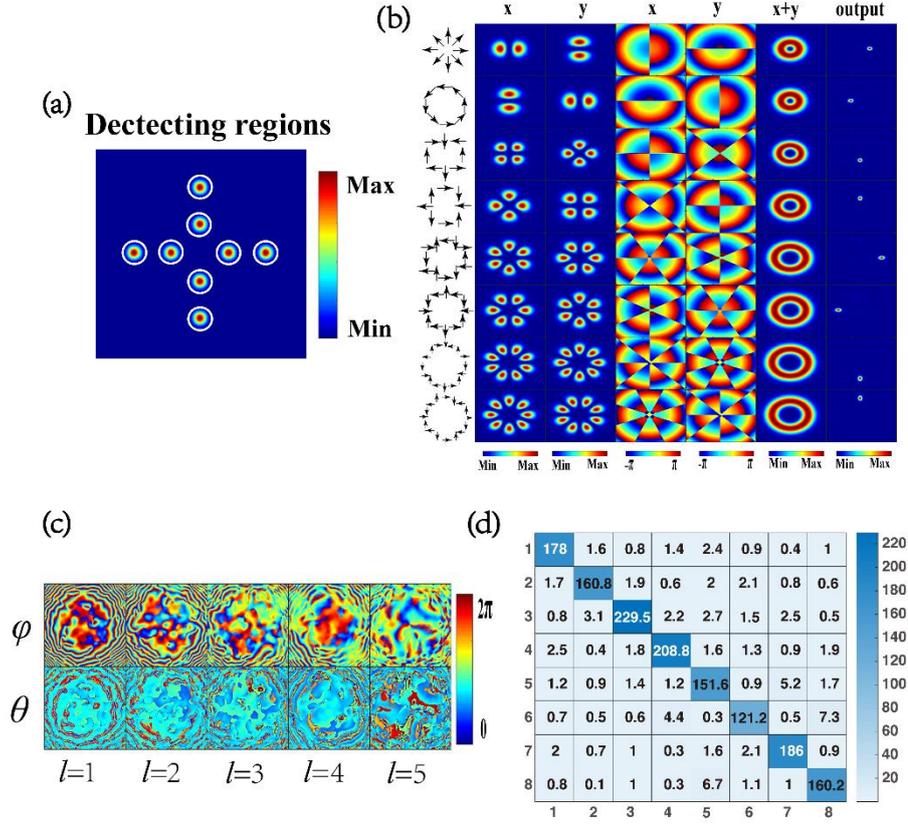

Figure 7. Classifying cylinder vector beams with different polarization states and topological charges ($l = \pm 1, \pm 2, \pm 3$ and $\pm 4$). (a) The output plane with 8 different classification regions. (b) Input cylinder vector beams with different topological charges and the results of classifier. (c)The optimized rotation angle and phase distributions. (d) The heat map of signal to noise of detecting results of classifying cylinder vector beams with different topological charges.

### 3.4 Generation and multiplexing of polarized OAM beams

On the other hand, we also try to use the PD$^2$NN to generate cylinder vector beams from polarized Gauss beam located in different areas in the input plane. We firstly use four 45 degrees linear polarized Gauss beams to train our network with 5 hidden layers to generate four kinds of cylinder vector beams which are the interference results of vortex beams with $l = \pm 3, \pm 4, \pm 5$ and $\pm 6$ respectively. The intensity of x and y polarized input beams and the total intensity of x and y polarized input light are shown in the first, second and third columns of Fig. 8 (a) respectively. The results of intensity and phase distributions of generated cylinder vector beams are shown in the last five columns of Fig. 8 (a). In order to achieve generated results with higher quality, we reduce the number of generated cylinder vector beams to 2. Then we consider using two 45 degrees linear polarized Gauss beams to train our network with 5 hidden layers to generate two kinds of cylinder vector beams whose polarization components are the interference results of vortex beams with $l = \pm 5$ and $\pm 6$ respectively. The intensity distribution of input light and output results are shown in Fig. 8 (b). Apparently, the quality of generated two kinds of cylinder vector beams is better than that of generated four kinds of cylinder vector beams.

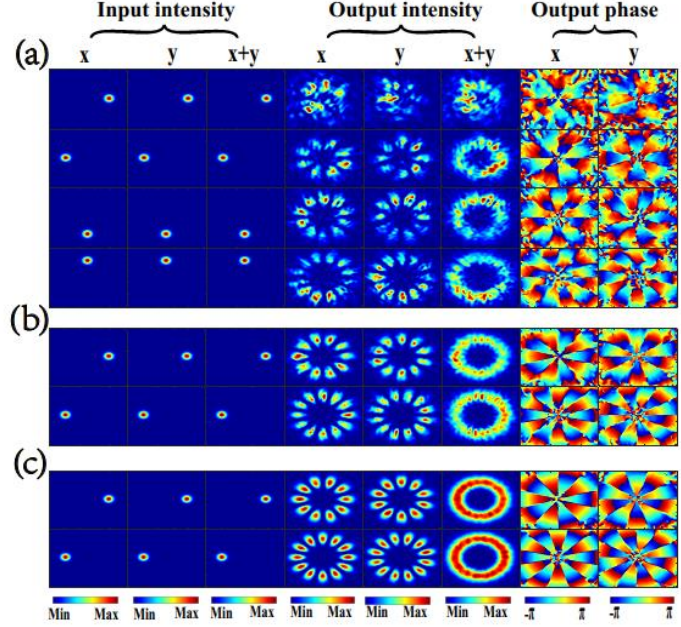

Figure 8 Generating cylinder vector beams. (a) Generating four kinds of cylinder vector beams (which are generated by the interference of vortex beams with $l = \pm3, \pm4, \pm5$ and $\pm6$ respectively) with diffractive layers number $N = 5$. (b) Generating two kinds of cylinder vector beams (which are generated by the interference of vortex beams with $l = \pm5$ and $\pm6$ respectively) with $N = 5$. (c) Generating two kinds of cylinder vector beams with $N = 10$.

In order to generate two kinds of cylinder vector beams with higher quality, we trained the network with more hidden diffractive layers. Fig. 8 (c) shows the results generated from a network with 10 layers have better quality than those from a network with 5 layers and are very close to the ground truth ones. For analyzing the effect of increasing the number of diffractive layers to generate two kinds of cylinder vector beams, we train the network with incremental layers from 3 to 10 and roughly use the mean square error (*MSE*) between the output intensity results and the ground truth ones to evaluate the generated modes quality by using the networks with different numbers of layers. The definition of intensity *MSE* can be expressed by:

$$MSE = \frac{1}{mn}\sum_{0}^{m-1}\sum_{0}^{n-1}[R(i,j) - T(i,j)]^2$$

where $R$ and $T$ represent the output intensity of PD$^2$NN and the ground truth intensity respectively and ($m$, $n$) represent the size of the light field. The results in Fig. 9 show that the *MSE* of intensity is roughly always on the downward trend with the number of diffractive layers increasing regardless of intensities in each polarization directions or total intensities in two polarization directions. But the results indicate that quality will not be enhanced very significantly as the number of layers increases when the number of diffractive layers is larger than 6. This process indicates that increasing the number of layers is an effective method to enhance the performance of our proposed network to generate cylinder vector beams. But it shall be noticed that increasing the number of hidden layers in D$^2$NN network can improve the quality of mode generation but also causes more loss of light power during the process of propagation between layers which is not favorable for increasing the output signal power.

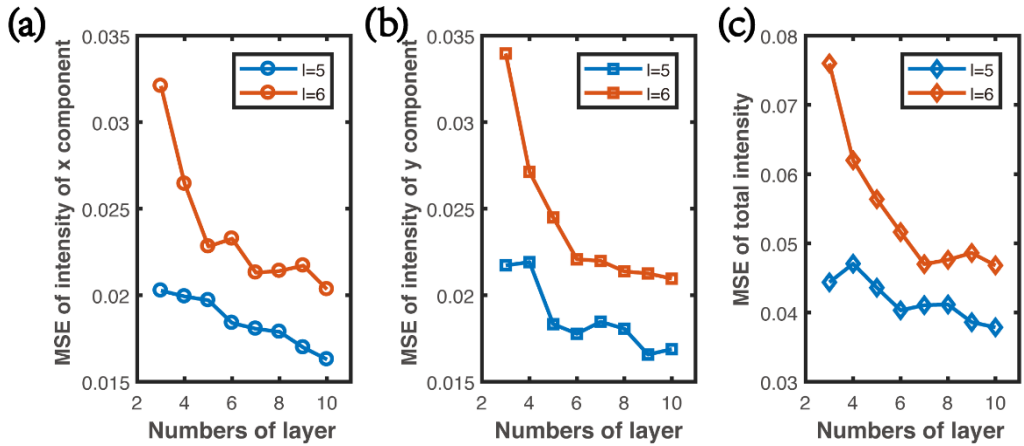

Figure 9 The intensity *MSE* results of generated two kinds of cylinder vector beam ($l = \pm5$ and $\pm6$) with different numbers of diffractive layers (a) *MSE* curve of intensity of x component, (b) *MSE* curve of intensity of y component, (c) *MSE* curve of total intensity with increasing layers from 3 to 10.

Additionally, this optical network trained with two 45 degrees linear beams to generate two cylinder vector beams (whose components are the interference of $l = \pm5$ and $\pm6$ respectively) also can be used to multiplex polarized linear beams at two locations in the input plane. The results of multiplexing are shown in Fig. 10 which are obtained from a trained network with 10 hidden diffractive layers. There are 8 kinds of input modes as input Gauss beams at two locations shown in first, second and third columns of Fig. 10 and 8 different cylinder vector beams generated shown in fourth to eighth columns of Fig. 10. The total intensity of every multiplexing result is different from each other which reflects the orthogonality. And the polarization state of each multiplexing result is different from each other. This multiplexing results indicate that this network can use linear polarized beams at two locations to generate a mixture of 8 kinds of hybrid cylinder vector beams.

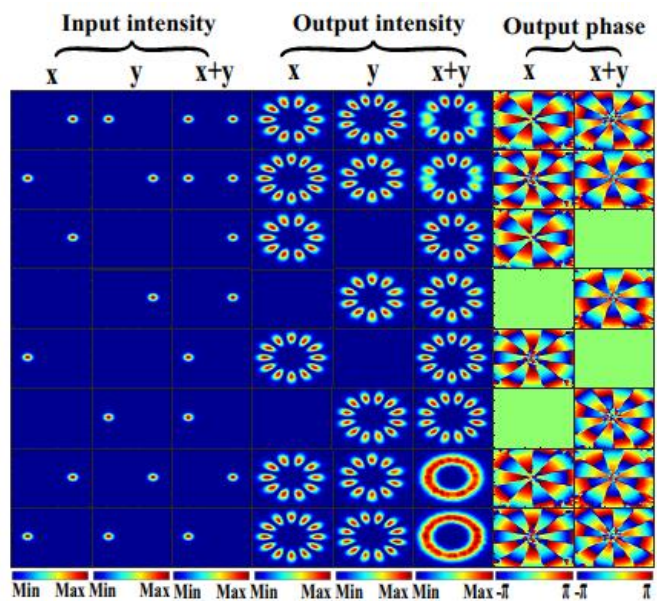

Figure 10 The de-multiplexing results of training two kinds of cylinder vector beam

# 4 Conclusion

In conclusion, we propose a novel phase-only modulated PD$^2$NN framework for processing polarized vortex beams and cylinder vector beams that conventional D$^2$NN cannot handle. It is implemented by designing proper rectangular micro-structures with wave-front matching algorithms. The hidden layers of the proposed network are determined by two parameters: modulated phase and rotation angle of designed rectangular micro-structure. The simulation results show that the proposed PD$^2$NN with 5 hidden layers can classify 14 kinds of polarized vortex beams with different topological charge numbers, 6 kinds of cylinder vector beams with identical topological charges and 8 kinds of cylinder vector beams into target Gauss beam spots in detecting regions. Additionally, the network with 10 layers can achieve generation and multiplexing of cylinder vector beams with a relatively high quality. Simulation shows that two linear Gauss beams can be successfully multiplexed into 8 kinds of cylinder vector beams. The simulation results also show that increasing the number of diffractive hidden layers of the network can improve the quality of generating and multiplexing hybrid OAM beams. This work lays the foundation for a further development of polarized neural network which can be applied in other potential fields such as multi-channel image recognition and logical operation of polarized light.

# 5 Reference

[1] J. Liu, P. Wang, X. Zhang, Y. He, X. Zhou, H. Ye, Y. Li, S. Xu, S. Chen, and D. Fan, Deep leaning based atmospheric turbulence compensation for orbital angular momentum beam distortion and communication, Optics Express 27, 16671-1688 (2019).
[2] F. Zhu, S. Huang, W. Shao, J. Zhang, M. Chen, W. Zhang, and J. Zeng, Free-space optical communication link using perfect vortex beams carrying orbital angular momentum (OAM). Optics Communications, 396, 50-57 (2017).
[3] M. Padgett and R. Bowman. Tweezers with a twist. Nature Photonics, 5,343-348 (2011).
[4] Y. Shen, X. Wang, Z. Xie, C. Min, X. Fu, Q. Liu, M. Gong, and X. Yuan. Optical vortices 30 years on: OAM manipulation from topological charge to multiple singularities. Light: science & applications, 8, 339-367 (2019).
[5] E. Nagali, F. Sciarrino, F. De Martini, L. Marrucci, B. Piccirillo, E. Karimi, and E. Santamato. Quantum information transfer from spin to orbital angular momentum of photons. Physical Review Letters, 103, 013601 (2009).
[6] G. B. Xavier, and G. Lima. Quantum information processing with space-division multiplexing optical fibres. Communications Physics, 3, 9, (2020).
[7] X. Fang, H. Ren, and M. Gu. Orbital angular momentum holography for high-security encryption. Nature Photonics, 14, 102-108 (2020).
[8] H. Zhou, B. Sain, Y. Wang, C. Schlickriede, R. Zhao, X. Zhang, Q. Wei, X. Li, L. Huang, and T. Zentgraf. Polarization-encrypted obital anular mmentum mltiplexed mtasurface holography. ACS Nano, 14, 5, 553-5559 (2020).
[9] S. Zheng, and J. Wang. Measuring Orbital Angular Momentum (OAM) States of Vortex Beams with Annular Gratings. Scientific Reports, 7, 40781, 1-9 (2017).
[10] R. Ryf, S. Randel, A. H. Gnauck, C. Bolle, A. Sierra, S. Mumtaz, M. Esmaeelpour, E. C. Burrows, R. J. Essiambre, and PJ Winzer. Mode-division multiplexing over 96 km of few-mode fiber using coherent 6 × 6 MIMO processing. Journal of Lightwave Technology, 30, 521-531 (2012).

[11] G. Gibson, J. Courtial, M. Padgett, M. Vasnetsov, V. Pasko, S. M. Barnett, and S. F. Arnold. Free-space information transfer using light beams carrying orbital angular momentum. Optics Express, 12, 5448-5456 (2004).

[12] C. Kai, P. Huang, F. Shen, H. Zhou, and Z. Guo. Orbital angular momentum shift keying based optical communication system. IEEE Photonics Journal, 9, 2 (2017).

[13] A. H. Gnauck, P. J.W. inzer, S. Chandrasekhar, X. Liu, B. Zhu, and D. W. Peckham. Spectrally efficient long-haul WDM transmission using 224-Gb/s polarization-multiplexed 16-QAM. Journal of Lightwave Technology, 29, 373-377 (2011).

[14] A. Sano, H.M asuda, T. Kobayashi, M. Fujiwara, K. Horikoshi, E. Yoshida, Y. Miyamoto, M. Matsui, M. Mizoguchi, H. Yamazaki, Y. Sakamaki, and H. Ishii. Ultra-high capacity WDM transmission using spectrally-efficient PDM 16-QAM modulation and C- and extended L-band wideband optical amplification. Journal of Lightwave Technology, 29, 578-586 (2011).

[15] X. Liu, S. Chandrasekhar, X. Chen, P. J. Winzer, Y. Pan, T. F. Taunay, B. Zhu, M. Fishteyn, M. F. Yan, J. M. Fini, E. M. Monberg, and F. V.D imarcello. 1.12-Tb/s 32-QAM-OFDM superchannel with 8.6-b/s/Hz intrachannel spectral efficiency and space-division multiplexed transmission with 60-b/s/Hz aggregate spectral efficiency. Optics Express, 19, B958-B964 (2011).

[16] S. Li, and J. Wang. A compact trench-assisted multi-orbital-angular-momentum multi-ring fiber for ultrahigh-density space-division multiplexing (19 rings × 22 modes). Scientific Reports, 4, 3853 (2014).

[17] X. Zhou, J. J. Yu, M. F. Huang, Y. Shao, T. Wang, L. Nelson, P. Magill, M. Birk, P. I. Borel, D. W. Peckham, R. Lingle, and B. Y. Zhu. 64-Tb/s, 8 b/s/Hz, PDM-36QAM transmission over 320 km using both pre- and post-transmission digital signal processing. Journal of Lightwave Technology, 29, 571-577 (2011).

[18] H. Huang, G. Milione, M. P. J. Lavery, G. Xie, Y. Ren, Y. Cao, N. Ahmed, T. A. Nguyen, D. A. Nolan, M. Li, M. Tur, R. R. Alfano, and A. E. Willner Mode division multiplexing using an orbital angular momentum mode sorter and MIMO-DSP over a graded-index few-mode optical fibre. Scientific Reports, 5, 14931, (2015).

[19] A. Wang, L. Zhu, J. Liu, C. Du, Q. Mo, and J. Wang. Demonstration of hybrid orbital angular momentum multiplexing and time-division multiplexing passive optical network. Optics Express, 23, 23, 29457-28466 (2015).

[20] Y. Fang, J. Yu, N. Chi, J. Zhang, and J. Xiao. A Novel PON Architecture Based on OAM Multiplexing for Efficient Bandwidth Utilization. IEEE Photonics Journal, 7(1):1-6 (2015).

[21] X. Hui, S. Zheng, Y. Hu, C. Xu, X. Jin, H. Chi, and X. Zhang. Ultralow reflectivity spiral phase plate for generation of millimeter-wave OAM beam. IEEE Antennas and Wireless Propagation Letters, 14, 966–969 (2015).

[22] V. Kotlyar, A. Kovalev, A. Porfirev, and E. Kozlova. Orbital angular momentum of a laser beam behind an off-axis spiral phase plate. Optics Letters, 44,15, 3673-3676 (2019).

[23] S. Slussarenko, A. Murauski, T. Du, V. Chigrinov, L. Marrucci, and E. Santamato. Tunable liquid crystal q-plates with arbitrary topological charge. Optics Express, 19, 5, 4085-4090 (2011).

[24] T. Arikawa, T. Hiraoka, S. Morimoto, F. Blanchard, and K. Tanaka. Transfer of orbital angular momentum of light to plasmonic excitations in metamaterials. Science Advanced, 6, 24 (2020).

[25] Y. Li, X. Li, L. Chen, M. Pu, J. Jin, M. Hong, and X. Luo. Orbital angular momentum multiplexing and demultiplexing by a single metasurface. Advanced Optical Materials, 5, 2, 1600502 (2017)

[26] Z. Huang, P.Wang, J. Liu, W. Xiong, Y. He, J. Xiao, H. Ye, Y. Li, S. Chen, and D. Fan. All-optical signal processing of vortex beams with diffractive deep neural networks. Physical Review Applied, 15, 014037 (2017).

[27] P. Wang, W. Xiong, Z. Huang, Y. He, J. Liu, H. Ye, J. Xiao, Y. Li, D. Fan, and S. Chen. Diffractive Deep Neural Network for Optical Orbital Angular Momentum Multiplexing and Demultiplexing," IEEE Journal of Selected Topics in Quantum Electronics, 28, 4, 1-112022 (2022).

[28] X. L. Wang, J. Ding, W. Ni, C. Guo, and H. Wang. Generation of arbitrary vector beams with a spatial light modulator and a common path interferometric arrangement. Optics Letters, 32, 24, 3549-3551 (2007).

[29] J. Zhang, Y. Xie, Q. Wu, and Y. Xia. Medical image classification using synergic deep learning. Medical Image Analysis, 54: 10-19 (2019).

[30] K. Noda, Y. Yamaguchi, K. Nakadai, H. G. Okuno, and T. Ogata. Audio-visual speech recognition using deep learning. Applied Intelligence, 42(4): 722-737 (2015).

[31] Y. Han, J. Jang, E. Cha, J. Lee, H. Chung, M. Jeong, T. Kim, B. G. Chae, H. G. Kim, S. Jun, S. Hwang, E. Lee, and J. C. Ye. Deep learning STEM-EDX tomography of nanocrystals. Nature Machine Intelligence, 3, 267-274 (2021).

[32] Y. LeCun, Y. Bengio, and G. Hinton. Deep learning. Nature, 521, 7553 (2015).

[33] X. Sui, Q. Wu, J. Liu, Q. Chen, and G. Gu. A review of optical neural networks. IEEE Access, 8, 70773-70783 (2020).

[34] X. Lin, Y. Rivenson, N. T. Yardimci, M. Veli, M. Jarrahi, and A. Ozcan. All-optical machine learning using diffractive deep neural networks. Science, 361, 6406, 1004-1008 (2018).

[35] L. Li, L. Zhu, Q. Zhang, B. Zhu, Q. Yao, M. Yu, H. Niu, M. Dong, G. Zhong, and Z. Zhou. Miniaturized Diffraction Grating Design and Processing for Deep Neural Network. IEEE Photonics Technology Letters, 31, 1952-1955 (2019).

[36] H. Chen, J. Feng, M. Jiang, Y. Wang, J. Lin, J. Tan, and P. Jin. Diffractive deep neural networks at visible wavelengths. Engineering, 7, 1483-1491 (2021).

[37] C. Qian, X. Lin, X. Lin, J. Xu, Y. Sun, E. Li, B. Zhang, and H. Chen. Performing optical logic operations by a diffractive neural network. Light: science & application, 9, 59 (2020).

[38]S. Watanabe, T. Shimobaba, T. Kakue and T. Ito. Hyperparameter tuning of optical neural network classifiers for high-order Gaussian beams. Optics Express, 30, 11079-11089 (2022).

[39] Y. Zuo, B. Li, Y. Zhao, Y. Jiang, Y. Chen, P. Chen, G. Jo, J. Liu, and S. Du. All-optical neural network with nonlinear activation functions. Optica, 6, 9 (2019).

[40] C. Liu, Qi. Ma, Z. Luo, Q. Hong, Q. Xiao, H. Zhang, L. Miao, W. Yu, Q. Cheng, L. Li, and T. Cui. A programmable diffractive deep neural network based on a digital-coding metasurface array. Nature Electronics, 5, 1-10 (2022).

[41] J. Zi, Y. Li, X. Feng, Q. Xu, H. Liu, X. Zhang, J. Han, and W. Zhang. All-function terahertz waveplate based on all-dielectric metamaterial. Physical Review Applied, 13, 034042 (2020).

[42] T. Wu, X. Zhang, Q. Xu, E. Plum, K. Chen, Y. Xu, Y. Lu, H. Zhang, Z. Zhang, X. Chen, G. Ren, L. Niu, Z. Tian, J. Han, and W. Zhang. Advanced optical materials, 10, 1 (2022).

[43] Y. Gao, S. Jiao, J. Fang, T. Lei, Z. Xie, and X. Yuan. Multiple-image encryption and hiding with an optical diffractive neural network. Optics Communications, 463, 125476 (2020).